# Cross-spectrum noise measurements at 4 K to minimize power-splitter anti-correlation effect

Archita Hati [a)], Craig W. Nelson, David Pappas, and David A. Howe

National Institute of Standards and Technology, 325 Broadway, Boulder, Colorado 80305, USA
[a)]E mail: archita.hati@nist.gov

**Abstract:** We report an accurate measurement of the phase noise of a thermally limited electronic oscillator at 300 K. By thermally limited we mean that the white signal-to-noise ratio of the oscillator is at or near the level generated by the thermal noise of the 50 Ω source resistor. The measurement is devoid of the anti-correlation effect that originates from the common mode power splitter in a cross-spectrum technique. The anti-correlation effect is mitigated by cooling the power splitter to a liquid helium temperature (4 K). The measurements in this paper are the first proof of theoretical claims that additive thermal noise from the splitter can be reduced significantly with cryogenic cooling and this can eliminate any anti-correlated noise introduced by use of the two-channel cross-spectrum technique. We also confirm measurements of partial anti-correlation error of (-1.3 ± 0.6) dB that agree with theory when the splitter is at liquid nitrogen temperature of 77 K.

## I. Introduction

One method of measuring the phase and amplitude fluctuations of an oscillator (referred to as the Device Under Test - DUT) is to measure its deviations relative to a reference phase and amplitude. To be flexible, many schemes can be configured for any given oscillator frequency, with the goal of keeping the measurement noise well below that introduced by the phase and amplitude detection system. The measurement-system noise of an optimized single-channel detection scheme very often exceeds the DUT intrinsic noise. For many years, phase-noise metrologists and instrument manufacturers have utilized a presumed powerful and widely used strategy to reduce this noise, namely, splitting the DUT signal to two identically optimized channels and computing the detected cross-spectrum using a two-channel spectrum analyzer [1]–[5]. This method significantly rejects each channel's noise by time-averaging the uncorrelated noise between each channel. Until recently, the deleterious effect of the intrinsic thermal noise added by the common mode power splitter on this measurement technique was unknown [6]–[8]. The thermal noise (or white spectrum noise) fluctuations originating from the Wilkinson power splitter (WPS) circuit is anti-correlated between its two outputs. This can produce an indeterminate error in the cross spectrum or may even grossly underestimate the white level, called cross-spectrum collapse [9], [10], if the DUT itself has thermally limited white noise [7]. A single-channel system is simply incapable of measuring a thermal-noise limited DUT; therefore, a two-channel cross-spectrum measurement system is necessary.

In our earlier work [7], different types of resistive power splitters, conventional and modified WPS were examined. However, they all demonstrated either positive or negative correlations, thus preventing an accurate two-channel cross-spectrum measurement near the thermal noise limit. One solution to this problem of anti-correlated thermal noise of the power splitter is briefly discussed in [11]. This technique uses a cross-correlation interferometer based on the WPS modified by removing the isolating resistor and rotating the undesired noise to the imaginary axis of the cross-spectrum while keeping the desired DUT noise on the real axis. The rotation is achieved by adding a controlled amount of equal delay at each output port of the power splitter. Maintaining the adjustment of the correct amount of delay over varying operating conditions is challenging and more research is in progress.

In this paper, we study and report on a cross-spectrum measurement that successfully reduces the effect of splitter anti-correlated noise by operating the splitter at low temperature.

## II. Cross-spectrum Phase Noise Measurement System at 4 K

Fig. 1 shows a schematic diagram of the phase noise measurement system. A pair of phase-sensitive detectors operate simultaneously to comprise the basic cross-spectrum phase noise measurement. The DUT signal goes through an impedance matching and harmonic filtering (IMHF), a calibrated additive white phase noise source (AWNS), and then to a Wilkinson 2-output power splitter (WPS) that is mounted inside of a cryostat. Each output of the WPS feeds separate, optimized single-channel phase detectors measuring the relative phase fluctuations between the DUT and separate oscillators (Ref #1 and Ref #2) using phase locked tracking loops (PLL) to maintain phase quadrature at the doubled balanced mixers (DBM1 and DBM2). The phase fluctuations at the mixer inputs convert to output voltage fluctuations at baseband frequencies ($f$) above the PLL bandwidth. The two-channel Fast Fourier Transform (FFT) analyzer then computes the cross-power spectral density ($S_{yx}$) between inputs $x$ and $y$ of channel 1 and 2. The AWNS is used to determine the mixer sensitivity. Also, a variable dc offset voltage (not shown) is added at the input of the PLL integrator to reduce the system sensitivity to DUT amplitude modulation (AM) noise [7].

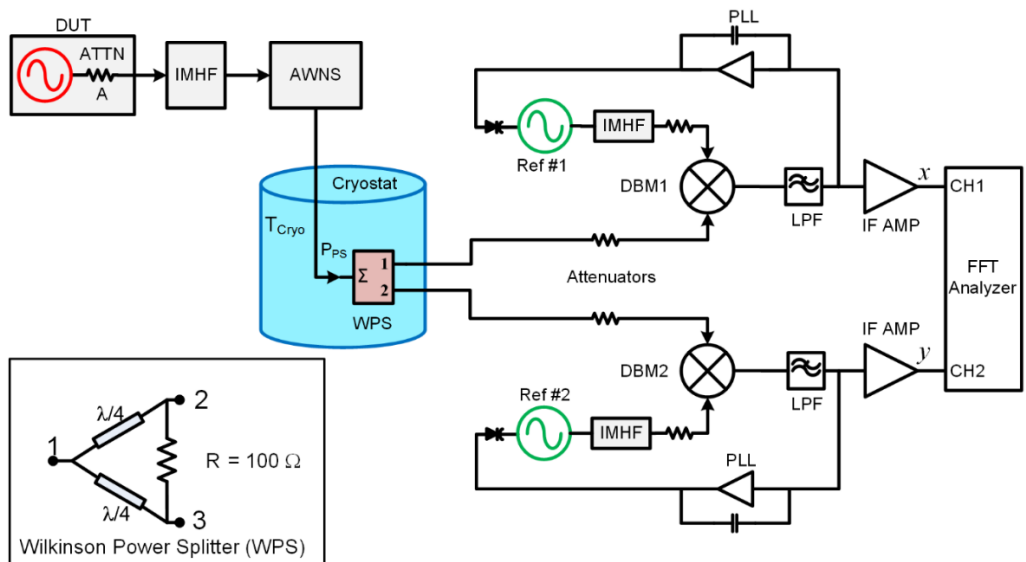

Fig. 1: Block diagram of a conventional dual-channel cross-spectrum system used for measuring phase noise of the DUT. The FFT analyzer measures the correlated spectrum which consists of the phase noise of the DUT, the anti-correlated noise of the Wilkinson power splitter (WPS) and the uncorrelated noise from separate detector channels which averages down toward zero, lowering the measurement system's overall noise floor. DBM – Double balanced mixer, LPF – Low Pass Filter, IMHF – Impedance Matching and Harmonic Filtering, AWNS – Additive White Noise Source, IF AMP – Intermediate Frequency (baseband) amplifier.

The measurements were taken in a variable temperature cryogenic system. The system has a liquid helium dewar and feedback controlled sample temperature from 1.8 to 350 K with stability of ±0.01 K. The sample space is shielded in a superconducting environment and is configured with three stainless steel RF coaxial cables. These cables were connectorized with SMA plugs to allow frequencies up to 26 GHz. The power splitter used for this experiment is shown in Fig. 2(a) and the sample fixture is shown in Fig. 2(b). Shown on the right (Fig. 2(c)) is the physical apparatus of the basic cross-spectrum phase noise measurement as shown in Fig. 1.

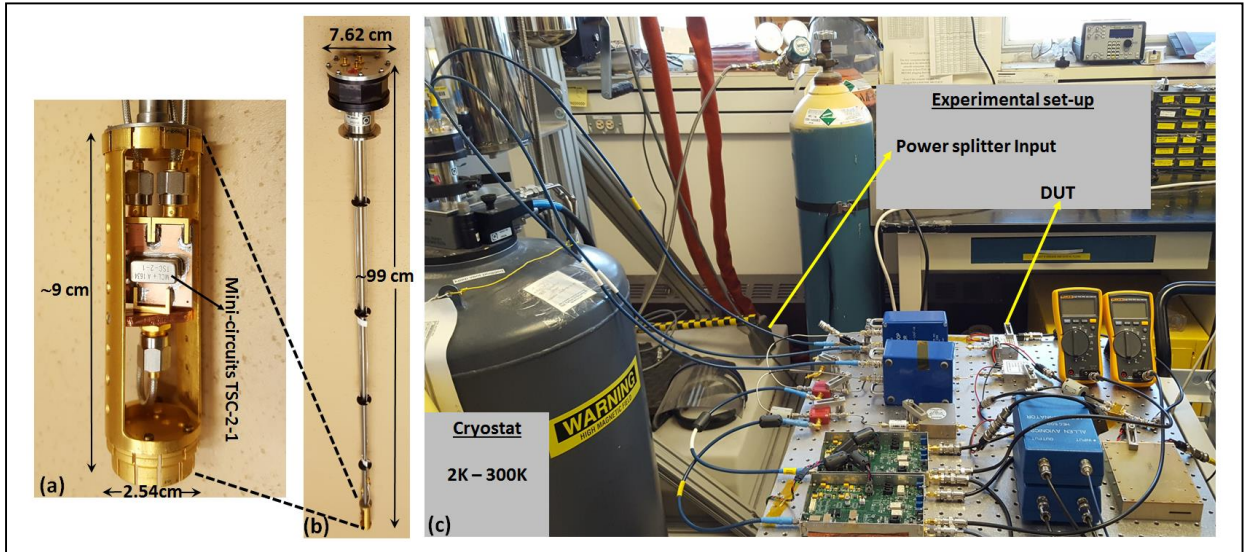

Fig. 2:(a) Picture of connectorized commercial-off-the-shelf (COTS) power-splitter that is carried in the mounting fixture which is inserted into the 4 K cryocooler. (b) Mounting fixture and (c) Experimental set-up used for the measurement of phase noise of the DUT.

The measured double sideband (DSB) phase noise $S_\varphi$, as expressed in (1), contains the desired DUT noise correlated in both measurement channels, the anti-correlated thermal noise of the power splitter, and also the uncorrelated instrument noise. The instrument noise approaches zero as $1/\sqrt{m}$, where $m$ is the number of FFT averages,

$$S_\varphi(f) = S_\varphi(f)\Big|_{DUT}^{T_{Elec}} - S_\varphi(f)\Big|_{WPS}^{T_{Cryo}} + \frac{S_\varphi(f)}{\sqrt{m}}\Big|_{Instrument}^{T_{Elec}} . \qquad (1)$$

Here, $T_{Cryo}$ and $T_{Elec}$ indicate the temperature of the cryostat and electronics, respectively. In a two-channel system like Fig. 1, as already discussed in [7], the WPS noise is anti-correlated (phase inverted) between channels 1 and 2 whereas the DUT noise is positively-correlated. At $T_{Cryo} = T_{Elec}$, the DUT and the WPS have equal thermal noise contributions which tend to annihilate each other in (1). This leads to cross-spectrum collapse and under this condition we measure only the instrument noise, i.e., the noise floor given by

$$S_\varphi(f) = \frac{S_\varphi(f)}{\sqrt{m}}\Big|_{Instrument}^{T_{Elec}} , \quad T_{Elec} = T_{Cryo} . \qquad (2)$$

However, when $T_{Cryo} \ll T_{Elec}$, the thermal noise of the WPS is significantly lower than the DUT thermal noise and essentially no cross-spectrum collapse occurs. The DUT noise can then be extracted and measured accurately for a sufficient number of FFT averages '$m$' as

$$S_\varphi(f) = S_\varphi(f)\Big|_{DUT}^{T_{Elec}} + \frac{S_\varphi(f)}{\sqrt{m}}\Big|_{Instrument}^{T_{Elec}} , \quad T_{Cryo} \ll T_{Elec}. \qquad (3)$$

When the WPS is kept at liquid Helium (4 K) and liquid Nitrogen (77 K) temperatures, its thermal noise is 18.8 dB and 5.9 dB, respectively, lower than the DUT thermal noise at 300 K. Therefore, the same measurement configuration as Fig. 1 should theoretically introduce an error of -0.1 dB and -1.3 dB respectively for these two temperatures.

Note that when both the DUT and the WPS are at same temperature, the anti-correlation effect dominates. By lowering the temperature of the WPS, an accurate measurement of the DUT noise is possible. However, when the DUT is at a lower temperature than the WPS, then anti-correlated noise of the WPS will be higher and the noise of the DUT cannot readily be measured.

## III.  Results

### A. *Characterization of the power splitter*

The Wilkinson power splitter for the measurement was chosen based on its good port-to-port isolation, low loss at 50 Ω terminating impedance, small size that fits the cryo-pump's fixture, and low-cost commercial-off-the-shelf (COTS) availability.  A suitable choice was the Mini-circuits TSC-2-1[1] and was not hand-selected in particular. Fig. 2a shows the placement of the power splitter on a small PC-board carrier. It was connectorized with three adaptors to interface SMA-fitted stainless steel semi-rigid coaxial cables. Relevant s-parameters (transmission and isolation) were measured as a function of coarse high temperature with fine-temperature measurements near 4 K. Fig. 3 shows that the transmission between port 1 to port 2 ($S_{21}$) was almost constant over the full temperature range.  Results for $S_{31}$ were similar to $S_{21}$. Isolation of the outputs was measured by the power transfer between output ports 2 and 3 ($S_{23}$).  $S_{23}$ isolation assures that the equivalent input noise of the mixer in one channel ("DBM" in Fig. 1) does not couple into the other channel enough to bias the measurement result.  Fig. 3 shows that the output-port isolation remained more than adequate, degrading only by about 7 to 8 dB and only at temperatures below 10 K. The transmission between input and outputs ports of the power splitter without the stainless steel coaxial cables was -3.3 dB. The S-parameters shown in Fig. 3 includes the cables loss of approximately 1.1 dB.

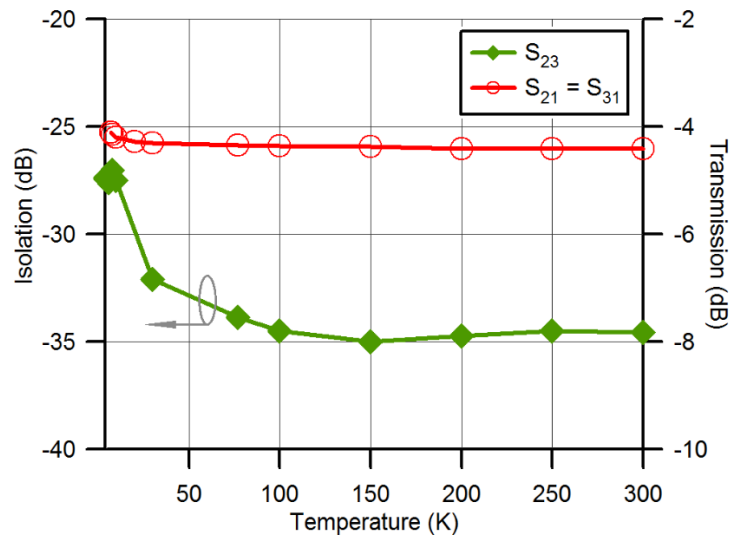

Fig. 3: Variation of isolation ($S_{23}$) and transmission ($S_{21}$ and $S_{31}$) of the power-splitter with temperature at 100 MHz.

### B. *Phase noise results*

Measurements were carried out on a commercial ultra-low noise quartz crystal oscillator operating at 100 MHz as the DUT [7], [12].  The principle and design of such oscillators for ultra-low thermal noise is described in [8], [13]. The DUT signal includes a resistive attenuator 'A' at the oscillator's output to assure a nominal 50 Ω output impedance match. At first phase noise of the DUT was measured with attenuator 'A' equal to 3 dB at the output of the oscillator. The net power '$P_{PS}$' into the power splitter (Fig. 1) was +10.5 dBm. Thus, the expected thermal-limited DSB phase noise of the DUT is given by the difference of wideband (Johnson) thermal noise to carrier power, or -174 dBrad$^2$/Hz – 10.5 dB = -184.5 dBrad$^2$/Hz. Referring to Fig. 4, the measured white noise level at 300 K using the cross-spectrum method is the gray plot that is predominantly well-below -184.5 dBrad$^2$/Hz.

---

[1] Commercial manufacturers are indicated for information purposes only. Other manufacturers exist.  No endorsement is implied.

Detection of such spectral collapse in the white noise region is difficult. However, in a case when two noise types intersect, it is easier to detect this problem. If a magnitude estimator $|<S_{yx}>|$ is used there will be an appearance of a notch in the magnitude of the cross-spectrum and if a $<\Re\{S_{yx}\}>$ estimator is used there will be a change in sign of the real part of the cross-spectrum [7]. An excellent analysis of different types of biased and unbiased estimators is described by Rubiola, et al. [8], [14]. All phase noise results presented in this paper are obtained from an unbiased $<\Re\{S_{yx}\}>$ estimator. In Fig. 4, the graph in inset (i) shows the real value of the cross-spectrum at 300 K where the phase is fluctuating around a mean of zero. A logarithmic plot of the same data with negative values omitted is shown in the gray curve of the main plot of Fig. 4.

When the splitter was cooled to 4 K, the anti-correlation features of the phase noise disappeared as shown by the yellow curve in Fig. 4. The average noise level is measured at the expected 50 Ω thermal noise from the DUT at 300 K. This measurement and those to follow are the first proof of theoretical claims that additive thermal noise from the splitter can be reduced, thus eliminating any anti-correlated splitter noise by use of the two-channel cross-spectrum technique.

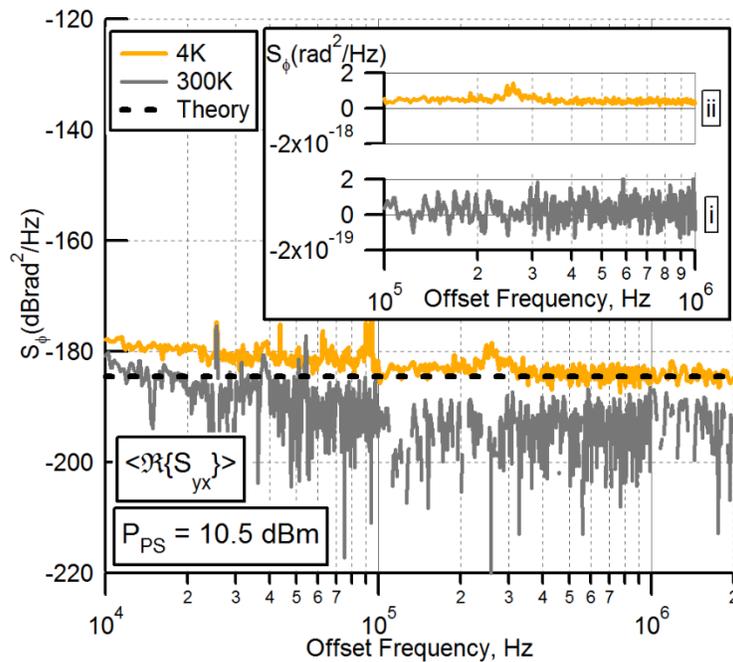

Fig. 4: Phase noise of a 100 MHz oscillator measured with a Wilkinson power splitter at 300 K and 4 K. Theoretical noise of this oscillator referenced to the input power of common-mode power splitter ($P_{PS}$) is respectively –184.5 dBc/Hz. The thermal noise is calculated from (–177 – $P_{PS}$). The measured noise at 4 K matches the theoretical value; however, there is a clear indication of spectrum collapse at 300 K as expected. Number of FFT averages '$m$' for 300 K and 4 K are 20,000 and 10,000 respectively for each frequency span.

Fig. 5 shows phase noise measurements at four different splitter temperatures in the range from 4 K to 77 K showing the change in the anti-correlation observations and to what degree it diminishes over this temperature range. The corresponding anti-correlation effect and full collapse at 300 K is also shown. We measured the expected partial collapse of $-1.3 \pm 0.6$ (1 σ) dB at 77 K as the observable difference between 4 K (yellow plot) and 77 K (purple plot) in the inset of Fig. 5. We see an anomalous noise artifact, a broadband noise around 260 kHz. This occurs only at low temperatures and disappears at 300 K; the cause is unknown but it does not affect the findings of this paper.

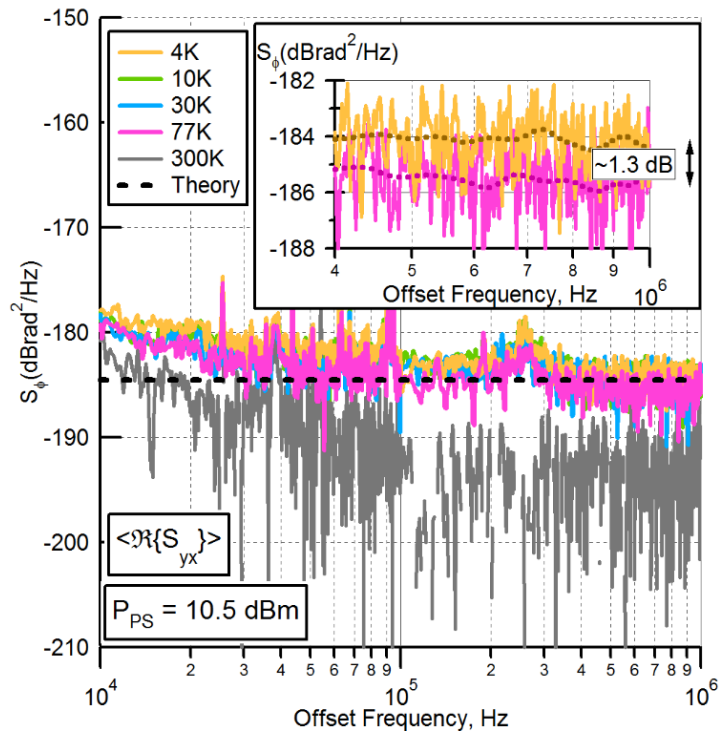

Fig. 5: Phase noise of a 100 MHz oscillator measured with a Wilkinson power splitter (WPS) at different cryogenic temperatures. Full collapse at 300 K and partial collapse (-1.3 ± 0.6) dB is observed at 77 K as expected. Below 30 K accurate measurements were observed. Number of FTT averages '$m$' used for 300 K is 20,000 and for temperatures 4K to 77 K is 10,000.

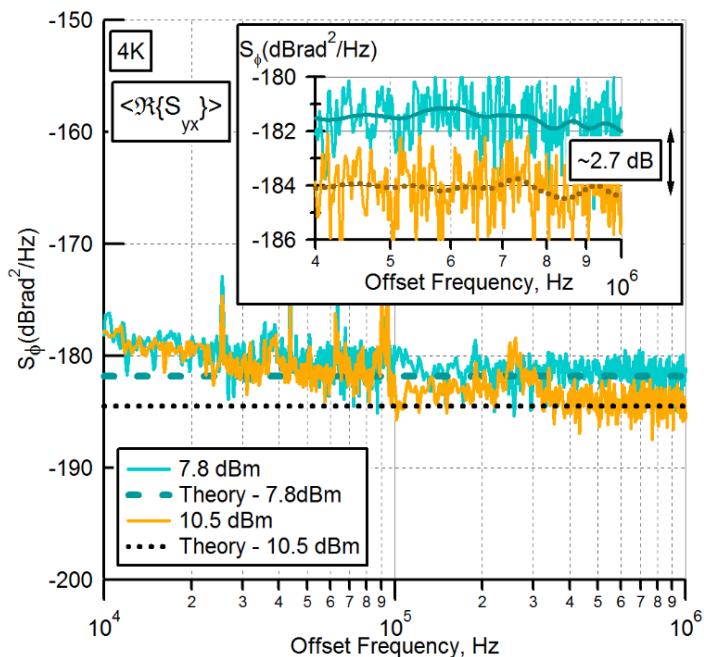

Fig. 6: Phase noise of a 100 MHz oscillator measured with a Wilkinson power splitter (WPS) at 4 K for different $P_{PS}$. The theoretical value for +7.8 dBm and 10.5 dBm are -181.8 and -184.5 dBrad$^2$/Hz respectively. The inset shows the difference between two thermal noise levels is nearly 2.7 ± 0.6 dB obtained from smoothing the individual curves by 1000 points.

To further verify the theory that the thermal noise is equal to (–174 –$P_{PS}$) dBrad$^2$/Hz we changed the input power of the power splitter. In [15], we showed that at 300 K we always get spectral collapse and it is independent of $P_{PS}$, the WPS input power. However, at 4 K we were able to measure the thermal noise of the DUT for different $P_{PS}$ and it matches the theory. Fig. 6 shows measured noise of the DUT at 7.8 dBm and 10.5 dBm that agrees with the theoretical value of - 181.8 dBrad$^2$/Hz and -184.5 dBrad$^2$/Hz respectively. The inset indicates that the difference between two thermal noise levels is nearly 2.7 ± 0.6 (1 σ) dB obtained by smoothing the individual curves.

## IV. Summary

In order to make measurements with a sufficiently low measurement noise floor and with a modest accuracy goal of ±1 dB, use of the cross-spectrum method and of an accurate calibration procedure are essential. The advantage of the cross-spectrum technique is that by splitting a signal between two measurement channels, noise from components in the noise measurement system is uncorrelated between the two channels, such that sufficient averaging reveals the noise of the DUT. However, as the power splitter itself is common to both measurement channels, anti-correlation of its noise can introduce significant errors exceeding ±1 dB due to "collapse" of the cross-spectrum. This paper shows the first proof that the theoretically expected levels can be measured with essentially no collapse-error present by use of the cross-spectrum technique when a Wilkinson power splitter (WPS) is operating at 4 K. We further confirm measurements of partial collapse of -1.3 ± 0.6 (1 σ) dB error that agree with theory when the WPS is at liquid nitrogen temperature of 77 K.

## Acknowledgement

The authors thank Dr. Fred Walls, Dr. Franklyn Quinlan, and Corey Barnes for helpful comments on this manuscript. This work is a contribution of NIST, an agency of the U.S. government, and is not subject to copyright. Commercial products may be indicated in this document, no endorsement is implied.